\documentclass[final,
3p,
sort&compress,
square,
]{elsarticle}
\usepackage{amssymb, amsfonts, amsmath}
\usepackage{mathrsfs}
\usepackage{xcolor}
\usepackage{chngcntr} 

\usepackage{booktabs} 

\usepackage{charter}

\journal{XXXX}

\begin{document}
\small

\begin{frontmatter}

\title{\flushleft \LARGE \bf Solubilization kinetics determines the pulsatory dynamics \\of lipid vesicles exposed to surfactant}

\author[ucsd]{Morgan Chabanon\corref{cor1}}%
\ead{mchabanon@eng.ucsd.edu} \cortext[cor1]{Corresponding author}
\author[ucsd]{Padmini Rangamani\corref{cor1}}%
\ead{prangamani@eng.ucsd.edu} 

\address[ucsd]{Department of Mechanical and Aerospace Engineering, University of California San Diego, La Jolla, 92093, CA, USA.}

\begin{abstract}
\noindent 
We establish a biophysical model for the dynamics of lipid vesicles exposed to surfactants. The solubilization of the lipid membrane due to the insertion of surfactant molecules induces a reduction of membrane surface area at almost constant vesicle volume. This results in a rate-dependent increase of membrane tension and leads to the opening of a micron-sized pore. We show that solubilization kinetics due to surfactants can determine the regimes of pore dynamics: either the pores open and reseal within a second (short-lived pore), or the pore stays open up to a few minutes (long-lived pore). First, we validate our model with previously published experimental measurements of pore dynamics. Then, we investigate how the solubilization kinetics and membrane properties affect the dynamics of the pore and construct a phase diagram for short and long-lived pores. Finally, we examine the dynamics of sequential pore openings and show that cyclic short-lived pores occur at a period inversely proportional to the solubilization rate. By deriving a theoretical expression for the cycle period, we provide an analytic tool to measure the solubilization rate of lipid vesicles by surfactants. Our findings shed light on some fundamental biophysical mechanisms that allow simple cell-like structures to sustain their integrity against environmental stresses, and have the potential to aid the design of vesicle-based drug delivery systems.

\end{abstract}


\end{frontmatter}


\section{Introduction}

\begin{figure*}[tb]
\centering
\includegraphics[scale=1]{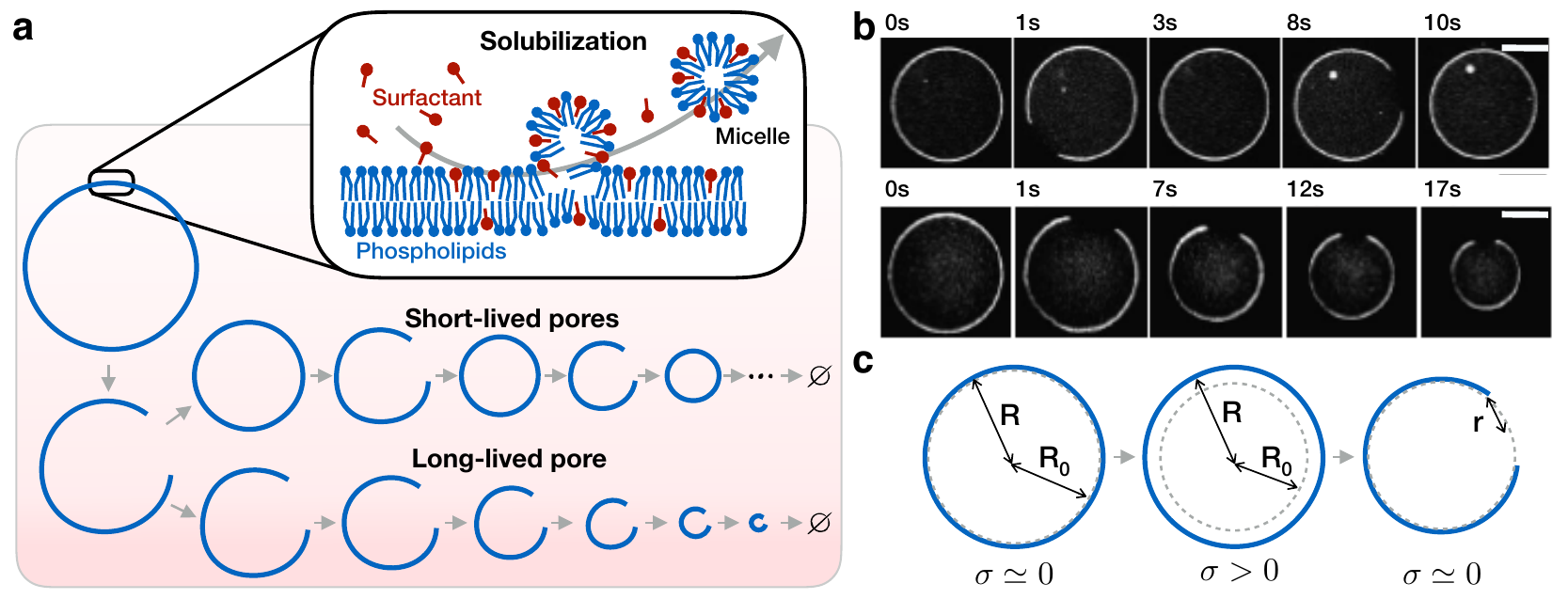}
\caption{Lipid vesicles exposed to surfactant exhibit two distinct behaviors -- either a succession of short-lived pores, or one long-lived pore followed by a series of short ones. (a) Schematic representation of the solubilization process of lipid vesicles by surfactants leading to either a succession of short-lived pores or one long-lived pore. (b) Microscopy images of DOPC vesicles exposed to surfactant TX-100 exhibiting a sequence of short-lived pores (top), and one long-lived pore (bottom). Scale bars are 10$\mu$m. (Adapted from \cite{hamada2009}). (c) Schematic representation of the area reduction mechanism leading to pore opening. Initially, the membrane is in low tension ($\sigma$) and the vesicle assumes a spherical shape of radius $R$ (left). Upon loss of surface area, the shrinking of the vesicle to its resting radius $R_0$ is frustrated by the encapsulated solution volume, leading to large membrane tension (middle). Once the rupture tension is reached, a circular pore of radius $r$ opens, releasing the membrane tension (right).}
  \label{fig:solu}
\end{figure*}


Surfactants, and more generally amphiphatic molecules, play important roles in many biological processes. For instance, lung surfactants are required for the surface area change of alveoli during breathing \cite{seifert2007, casals2012}, while bile salts facilitate fat absorption and interact with the bacteria flora in the small intestine and colon \cite{ridlon2016}. Biological surfactants, such as saponins secreted by plants, serve as defense mechanisms because of their ability to permeabilize lipid membranes and complex cholesterol \cite{rosenberg1999, francis2002}. Certain microorganisms produce surfactants to control the biochemical and biophysical properties of their surface, for example, by regulating the availability of water-insoluble molecules, or by modulating their adhesion properties \cite{rosenberg1999}. Antimicrobial peptides are amphiphatic molecules, whose actions are often compared to surfactants due to their propensity to insert into and permeabilize lipid bilayers \cite{ladokhin2001, henriksen2010}, although more specific mechanisms seem to be at play, such as the generation of negative Gaussian curvature \cite{schmidt2013}. Artificial and natural surfactants are also largely used in medical and biotechnological applications for their antimicrobial properties (see \cite{heerklotz2008} and references within), for isolation of membrane proteins \cite{lemaire2000}, and as permeabilizing agents to facilitate transport of drugs or DNA across cell membranes \cite{clamme2000, pierrat2013}. Thus, the interaction of surfactant molecules with lipid bilayers is central to many processes across the plant and animal kingdom.



One of the fundamental features of surfactant-membrane interactions is that the surfactants can insert themselves within the lipid bilayer and alter the surface area of the membrane through lipid solubilization (Fig.~\ref{fig:solu}(a)). The classical model describing the behavior of the surfactant-lipid systems as a function of the surfacant's relative concentration is the three-stage model proposed by Helenius and Simons \cite{helenius1975}: first, at low surfactant concentration, the surfactant molecules partition into the membrane; second, above a critical surfactant concentration, membrane solubilization occurs and mixed micelles coexist with the lipid membrane, and finally, above a second critical surfactant concentration, only micelles remain. It is important to note that this description is at equilibrium, and that in a lot of experimental settings, the surfactant concentration is large enough to induce micelle formation before the first stage reaches equilibrium \cite{lichtenberg2013a}, highlighting the need for out-of-equilibrium studies on membrane solubilization kinetics.

In this work, we focus on the effect of solubilization kinetics on the dynamic evolution of lipid vesicle morphology. 
Nomura \textit{et al.} \cite{nomura2001} investigated the time evolution of lipid vesicles exposed to various surfactants and observed several dynamic outcomes depending on the surfactant type and concentration: continuous shrinkage, cyclic shrinkage, minute-long pore opening, or inside-out inversions. More recently, this list of outcomes was extended by Hamada \textit{et al.} \cite{hamada2012}, and explained using a conceptual model for the different vesicle dynamics. Interestingly, the main observed outcomes were that spherical vesicles decrease in size and exhibit cyclic pore openings, with the first pore that was either short-lived ($\sim$ 1 second) or long-lived ($\sim$ 1 minute) \cite{hamada2009, hamada2012} (see Fig.~\ref{fig:solu}(a) and (b)). Although it has been observed that surfactant concentration and vesicle size play a role in determining if the vesicle will exhibit a short or long-lived first pore, a quantitative understanding of the membrane dynamics in the presence of surfactant is currently missing \cite{kaga2006, hamada2009}. In particular, the influence of physical factors such as pore line tension, membrane stretch modulus, and surfactant solubilization kinetics have not yet been fully investigated.




Here, we propose a quantitative mathematical model for the dynamics of a lipid vesicle that undergoes membrane area reduction due to exposure to a surfactant. We account for membrane solubilization by a rate of lipid removal through micelle formation. The resulting membrane area reduction produces an increase in intravesicular pressure, leading to the opening of a pore that is micrometers in size (Fig.~\ref{fig:solu}). This model captures both types of pore dynamics -- short-lived pores or a long-lived pore followed by short-lived ones. After validating our model by comparing its predictions with published experimental data, we conduct a systematic exploration of the influence of the physical parameters on the pore and vesicle dynamics. We show that the solubilization kinetics and the pore line tension are the dominant parameters controlling the dynamics of the pore. Finally, we demonstrate that the cycle period depends on the solubilization rate and derive an analytical expression that allows us to obtain this rate from experimental data.

\section{Model development}


\subsection{Model formulation} \label{eq:sec_model}

First, we propose a governing equation that describes the evolution of the membrane area as a function of the surfactant parameters. We assume that the primary phenomenon driving the vesicle area reduction is the production of mixed micelles due to the surfactant inserted in the membrane. This hypothesis is motivated by the fact that non-ionic surfactants have very fast insertion and flip-flop rates (e.g. $\sim$100~ms for Triton-X~100 \cite{alonso1987, heerklotz2003, tsamaloukas2006}) compared to the vesicle dynamics ($\sim$ minute \cite{hamada2009, hamada2012}), implying that surfactant saturation in the membrane occurs almost instantaneously. Consequently, we assume that the solubilization rate $k$ is determined by the micelle formation kinetics. Moreover, the membrane properties and the solubilization rate are fixed for a given total surfactant concentration. It follows that the time evolution of the reference, unstressed, membrane area $A_0$ can be written as
\begin{equation} \label{eq:dA0dt}
\frac{dA_0}{dt} = -k A_0 .
\end{equation}

We consider the total lipid membrane, with surface area $A_m$, to be elastic and define the membrane tension as
\begin{equation} \label{eq:sigma}
\sigma = \kappa \epsilon = \kappa \frac{A_m - A_0}{A_0} ,
\end{equation}
where $\kappa$ is the area stretch modulus and $\epsilon$ is the membrane strain.
While the reference membrane area $A_0$ decreases due to surfactant (Eq.~\eqref{eq:dA0dt}), the actual membrane area $A_m$ is constrained by the volume enclosed in the vesicle (see Fig.~\ref{fig:solu}(c)). This results in an increase in strain $\epsilon$ and membrane tension $\sigma$, until the bilayer eventually ruptures.

The membrane area, $A_m$, is constrained by its enclosed volume $V$, whose evolution is a function of the solvent fluxes out of the vesicle. In the absence of osmotic differential, the change of vesicle volume is the sum of the flux of solvent out the vesicle through the membrane, and through the pore  $dV/dt = J_m + J_p$. First, the flux of solvent permeating through the surface area of the lipid membrane $A_m$ is induced by the Laplace pressure due to membrane tension $\Delta p (\sigma)$, such that $J_m=-A_m P \Delta p(\sigma)$; here $P$ is a coefficient characterizing the permeability of the membrane to water defined as $P=P_s \nu_s /(k_B T N_A)$, where $P_s$ is the permeability of the membrane to solute, $\nu_s$ is the molar volume of the solvent, and $k_B$, $T$, and $N_A$  the Boltzmann constant, the absolute temperature of the system, and the Avogadro's number respectively. Second, the flux of solvent through the surface area of the pore $A_p$ is driven by a leak-out velocity $v(\sigma)$ induced by the membrane tension, leading to $J_p=-A_p v(\sigma)$.
Therefore we can write the vesicle volume dynamics as
\begin{equation} \label{eq:dVdt}
\frac{dV}{dt}= - [A_m P \Delta p(\sigma) +   A_p v(\sigma) ] .
\end{equation}

Finally, the dynamics of the pore circumference $L_p$  can be modeled as an over-damped system as 
\begin{equation} \label{eq:dLpdt}
\zeta \frac{dL_p}{dt} = F(\sigma, L_p) ,
\end{equation}
where $\zeta$ is the membrane drag coefficient, and $F(\sigma, L_p) $ is a conservative force arising from the elastic energy and pore energy. The membrane drag has two contributions $\zeta = \alpha_1 \eta_m + \alpha_2 \eta_s$, one from membrane dissipation, proportional to the lipid bilayer viscosity $\eta_m$,  and one from the membrane friction with the solvent, proportional to the solution viscosity $\eta_s$. Here $\alpha_1$ and $\alpha_2$ are geometric coefficients, of length dimensions, which we will specify later.

Equations \eqref{eq:dA0dt}, \eqref{eq:dVdt} and \eqref{eq:dLpdt} are the three governing equations of the system. However, we have five geometric variables ($V$, $A_0$, $A_m$, $A_p$, $L_p$). In order to reduce the number of variables, we assume the vesicle to be a sphere of radius $R$, with a circular pore of radius $r$ (see Fig.~\ref{fig:solu}(c)). Additionally, we define the radius of the reference area as $R_0$. It follows that 
\begin{equation}
\begin{array}{c}
V=4/3 \pi R^3 , \quad A_0 = 4 \pi R_0^2 , \quad A_m = 4\pi R^2 - \pi r^2 , \\
A_p = \pi r^2 , \text{ and} \quad L_p = 2 \pi r .
\end{array}
\end{equation}
Furthermore, the Laplace pressure in a spherical vesicle is $\Delta p (\sigma) = 2\sigma /R$, and the flow through a circular pore at low Reynolds number is $v(\sigma)=\Delta p(\sigma) r / (3 \pi \eta_s) = 2 \sigma r / (3 \pi \eta_s R)$ \cite{happel1983}. 

We can now write the conservative force as $F(\sigma,r)=-\partial V(\sigma,r)/\partial r$, where the membrane potential $V(\sigma,r)=V_s + V_p$ is the sum of the strain energy $V_s=\kappa (A_m-A_0)^2/(2A_0)$ and the pore energy $V_p = 2\pi r \gamma$, where $\gamma$ is the line tension of the pore. Noting that $\sigma = \partial V_s /\partial A$, the conservative force becomes $F(\sigma,r)=2\pi \sigma r - 2\pi \gamma$. Finally, the geometric drag coefficient for a circular pore in a membrane of thickness $h$ are \cite{ryham2011, aubin2016, ryham2018} $\alpha_1=h$ and $\alpha_2=2\pi r$. 

With these definitions, the three equations governing the vesicle dynamics take the form
\begin{equation} \label{eq:dR0dt}
2\frac{dR_0}{dt} = -k R_0 ,
\end{equation}
\begin{equation} \label{eq:dRdt}
4\pi R^2\frac{dR}{dt}= -\frac{2\sigma}{R} \left( A_m P  +   \frac{A_p r}{3\pi\eta_s} \right) ,
\end{equation}
and
\begin{equation}\label{eq:drdt}
\left(\eta_m h + 2\pi \eta_s r\right)\frac{dr}{dt} = \sigma r - \gamma .
\end{equation}

Finally, we define the pore nucleation mechanism. Following the classical nucleation theory, the energetic cost to open a pore of radius $r$ in a tense membrane can be computed based on the membrane potential $V(\sigma, r)-V(\sigma,0)$. The energetic cost to open a pore in a tense membrane presents a energy barrier at a critical pore radius $r_c(\sigma)$ that depends on the membrane tension \cite{koslov1984, idiart2004, chabanon2017}. The higher the membrane tension, the lower the critical pore radius and the corresponding energy barrier. It was shown that the stochastic nature of membrane thermal fluctuations helps overcome the energy barrier for the formation of a pore in a load rate-dependent manner \cite{chabanon2017}: a membrane stretched faster breaks at a higher tension on average \cite{evans2003, boucher2007, evans2011, bicout2012, chabanon2017}. Such a consideration is important if one wishes to capture the long time dynamics of vesicle undergoing multiple swell burst cycles \cite{chabanon2017, su2018}. In the present study, however, we focus on the pore dynamics of the few first pores, where the rate dependence of the rupture tension only weakly influences the system's behavior. Therefore, for simplicity, we will assume here that pore nucleation occurs at a constant critical tension. Accordingly, we prescribe a critical strain $\epsilon^*$, at which an initial pore large enough to overcome the nucleation barrier $r_0=\gamma/(\kappa \epsilon^*)$ is artificially created.

\subsection{Dimensionless system}

We begin by defining the following dimensionless variables
\begin{gather}
\bar{R} \equiv R/R_i , \quad \bar{R}_0 \equiv R_0/R_i , \quad  \bar{r}\equiv r / R_i ,
\end{gather}
where $R_i$ is the initial GUV radius. We further define the dimensionless time with respect with the characteristic time associated with the pore kinetics \cite{brochard-wyart2000}
\begin{equation}
\bar{t} \equiv t / \tau , \quad \text{with} \; \tau \equiv \eta_m h / \kappa .
\end{equation}
Introducing the above non-dimensional quantities in Eqs.~\eqref{eq:dR0dt}, \eqref{eq:dRdt} and \eqref{eq:drdt}, the system takes the dimensionless form
\begin{equation} \label{eq:dR0dt_dimless}
\frac{d \bar{R}_0}{d\bar{t}} = -\Theta \bar{R}_0 ,
\end{equation}
\begin{equation} \label{eq:dRdt_dimless}
\Lambda  \frac{d\bar{R}}{d\bar{t}} = -\frac{\epsilon}{\bar{R}^3}\left[ \Phi (4\pi \bar{R}^2 - \pi \bar{r}^2) + \bar{r}^3 \right] ,
\end{equation}
and
\begin{equation} \label{eq:drdt_dimless}
\left( 1+ \frac{\Lambda}{3} \bar{r} \right)\frac{d\bar{r}}{d\bar{t}}= \epsilon \bar{r} - \Gamma .
\end{equation}
Here, the non-dimensional parameters are defined as
\begin{equation} \label{eq:Lambda}
\Lambda \equiv \frac{6 \pi \eta_s R_i}{\eta_m h} = \frac{\text{effect of solution viscosity}}{\text{effect of membrane viscosity}}
\end{equation}
\begin{equation} 
\Phi \equiv \frac{3 \eta_s}{R_i}\frac{P_s \nu_s }{ k_B T N_A} = \frac{\text{flux through the membrane}}{\text{flux due to leakout}}
\end{equation}
\begin{equation} \label{eq:Gamma}
\Gamma \equiv \frac{\gamma}{\kappa R_i} = \frac{\text{pore line tension}}{\text{membrane surface tension}}
\end{equation}
\begin{equation} \label{eq:Theta}
\Theta \equiv \frac{\eta_m h}{\kappa} \frac{k}{2} = \frac{\text{pore time scale}}{\text{solubilization time scale}}
\end{equation}

Note that the analytic solution of Eq.~\eqref{eq:dR0dt_dimless} is 
\begin{equation} \label{eq:R0}
\bar{R}_0(\bar{t}) = e^{-\Theta \bar{t}} .
\end{equation}
Therefore we only need to solve Eqs.~\eqref{eq:dRdt_dimless} and \eqref{eq:drdt_dimless} numerically with Eq.~\eqref{eq:R0} as an input, and the pore nucleation mechanism described in Section~\ref{eq:sec_model}.

\subsection{Numerical implementation}

Numerical computations were carried using a custom code in MATLAB \textsuperscript{\textregistered}(Mathworks, Natick, MA) based on the code developed in\cite{chabanon2017}. The dimensionless constitutive equations~\eqref{eq:dRdt_dimless} and \eqref{eq:drdt_dimless}, coupled to Eq.~\eqref{eq:R0}, were solved using the Euler method, with a non-dimensional time step of 1 (smaller time steps did not improve the accuracy of the results significantly). All parameters values were as given in the figures, with initial dimensionless variables of $\bar{R}(\bar{t}=0)=1$ and $\bar{r}(\bar{t}=0)=0$. The pore nucleation mechanism was as follows: if the membrane area strain was greater or equal to the critical strain and no pore was open ($\epsilon\ge \epsilon^* \wedge \bar{r}=0$), a pore of radius $\bar{r}_\text{nuc}=\Gamma/\epsilon^*$ was nucleated. MATLAB\textsuperscript{\textregistered} codes are available upon request to the authors.


\section{Results}

\begin{figure}[tb]
\centering
\includegraphics[scale=1]{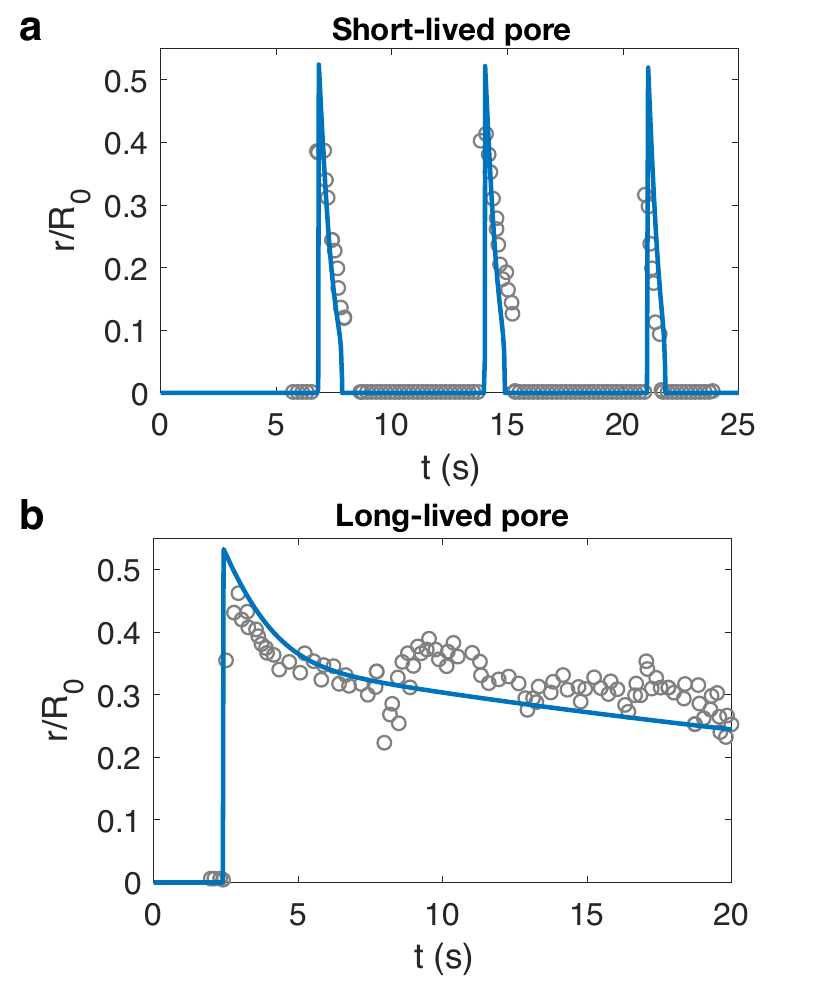}
\caption{Model predicts two regimes of pore dynamics in vesicles exposed to surfactants, in agreement with experimental observations: (a) Short-lived pores induced by low experimental concentration of surfactant, equivalent to slow solubilization kinetics in simulations ($k=1.4\times10^{-2}$ s$^{-1}$). (b) Long-lived pore observed at high experimental concentration of surfactant, corresponding to large solubilization kinetics in simulation ($k=4\times10^{-2}$ s$^{-1}$). Pore radii normalized by the vesicle reference radius as a function of time. Solid lines are model predictions, grey circles are experimental measurements of DOPC vesicles subject to TX-100 surfactant, corresponding to Fig.~\ref{fig:solu}(b). Experimental data are adapted from \cite{hamada2009}, with the initial time adjusted to match the first pore opening. Model parameters for this figure are reported in Table~\ref{tab:para}.}
\label{fig:comp}
\end{figure}

\begin{table}[tb]
\small
\center
\begin{tabular}{ l l }
\hline
 \multicolumn{2}{l}{ Physical parameters} \\
 $R_i$ & 10$\times 10^{-6}$ m \\
$\kappa$ & 2$\times 10^{-4}$ N/m \\ 
$\gamma$ & SLP: 1.2$\times 10^{-12}$ N ;~ LLP: 3$\times 10^{-13}$ N \\ 
$k$ & SLP: 1.4$\times 10^{-2}$ 1/s ;~ LLP: 4$\times 10^{-2}$ 1/s \\
$h$  & 7$\times 10^{-9}$ m \\ 
$\eta_m$ & 5 Pa s \\ 
$\eta_s$ & $10^{-3}$ Pa s\\ 
$\epsilon^*$ & 0.1 \\ 
$P_s$ & 2 $\times 10^{-5}$ m/s \\
$\nu_s$ & 18.04$\times 10^{-6}$ m$^3$/mol \\ 
$k_B$ & 1.38$\times 10^{-23}$ J/K\\ 
$N_A$ &  6.023$\times 10^{23}$ 1/mol \\ 
$T$ & 294 K \\
\hline
\multicolumn{2}{l}{ Dimensionless parameters} \\
$\Lambda$ &  5.39  \\
$\Phi$ & 4.43$\times 10^{-11}$\\
$\Gamma$ &  SLP: 6 $\times 10^{-4}$ ;~ LLP: 1.50$\times 10^{-4}$ \\
$\Theta$ & SLP: 1.23$\times 10^{-6}$ ;~ LLP: 3.50$\times 10^{-6}$ \\
\hline
\end{tabular} 
\caption{Values of the physical and corresponding dimensionless parameters used to compute the pore dynamics shown in Fig.~\ref{fig:comp}. SLP: Short-lived pore, LLP: Long-lived pore.}
\label{tab:para}
\end{table}


\begin{figure*}[tb]
\centering
\includegraphics[scale=1]{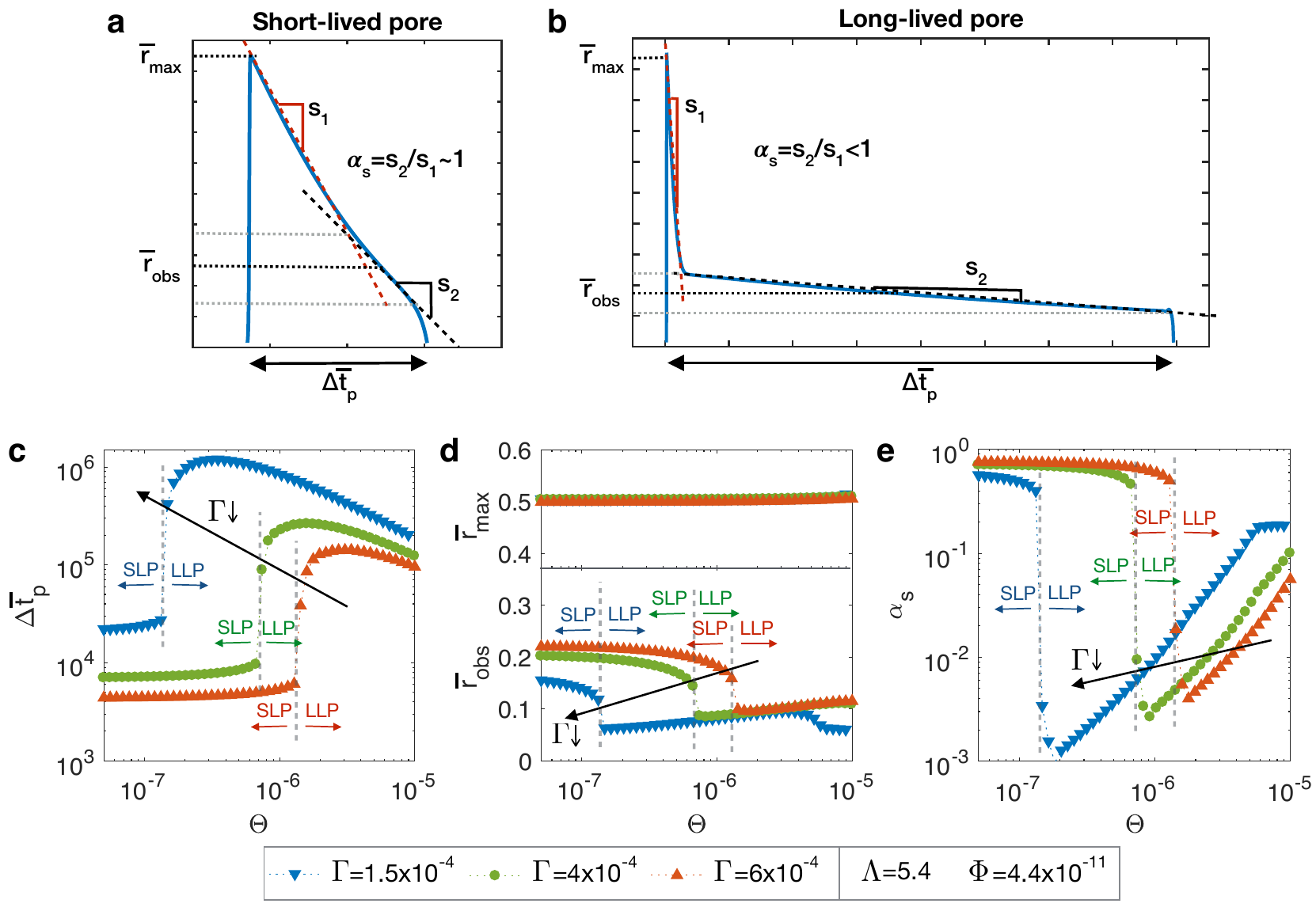}
\caption{Long-lived pores are facilitated by high solubilization rates and low values of $\Gamma$. 
(a, b) Definition of four metrics for short (a) and long-lived pores (b): the pore life time $\Delta \bar{t}_p$, the maximum pore radius $\bar{r}_\text{max}$, the observed pore radius $\bar{r}_\text{obs}$, and the slope ratio $\alpha_s$ between the slow phase and the fast closing phases (more details on the definitions of the metrics are given in the text and in Section~\ref{S_metrics} of the Supplementary Material).
(c-e) Influence of $\Theta$ and $\Gamma$ on the pore life time (c), maximum and observed radii (d), and slope ratio (e) for other parameters kept constant. Black arrows indicate decreasing values of $\Gamma$. Color arrows indicates the ranges of $\Theta$ were the vesicle exhibit a short-lived pore (SLP), or a long-lived pore (LLP). The transition between short and long-lived pores is abrupt, as seen from the discontinuity in the pore life time, observed radius, and slope ratio.}
\label{fig:MTheta}
\end{figure*}

\begin{figure*}[tb]
\centering
  \includegraphics[scale=1]{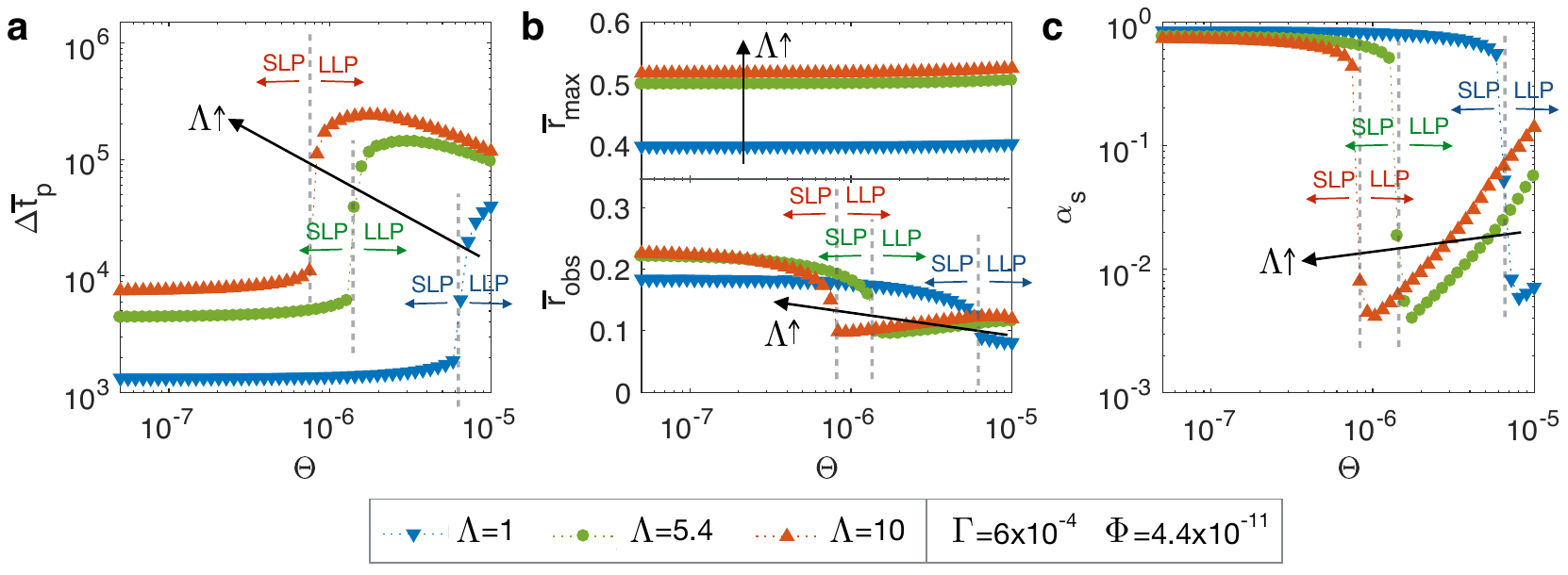}
  \caption{Higher values of $\Lambda$ lead to long-lived pores and smaller maximum pore radii. Influence of $\Theta$ and $\Lambda$ on the pore life time (a), maximum and observed radii (b), and slope ratio (c) for other parameters kept constant. Black arrows indicate increasing values of $\Lambda$. Color arrows indicates the ranges of $\Theta$ were the vesicle exhibit a short-lived pore (SLP), or a long-lived pore (LLP). }
  \label{fig:Lambda}
\end{figure*}

\begin{figure*}[t!b]
\centering
  \includegraphics[scale=1]{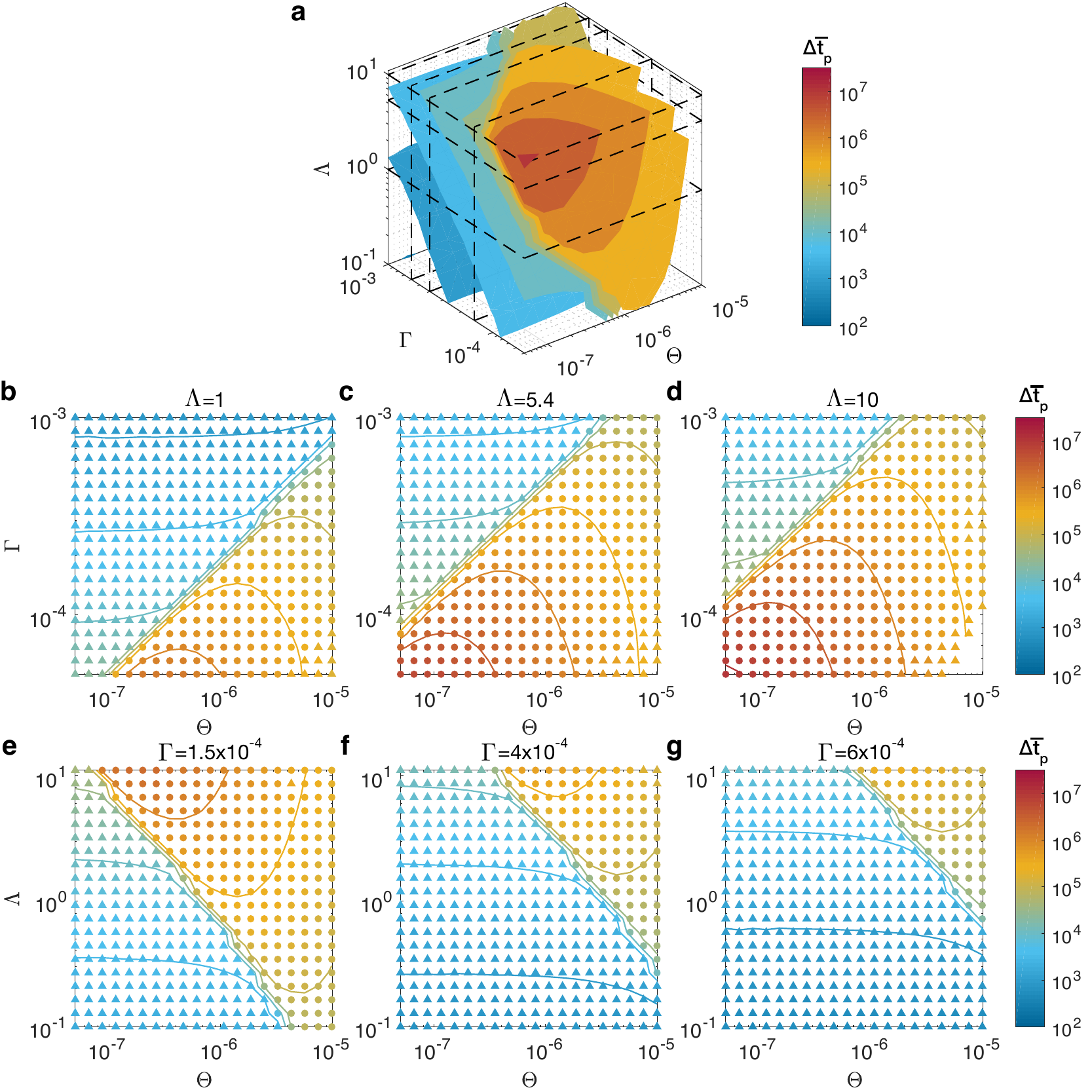}
  \caption{The transition between short and long-lived pores is a plane in the $(\Theta, \Gamma, \Lambda)$ parameter space. (a) Isocontours of the pore lifetime in the $(\Theta, \Gamma, \Lambda)$ parameter space. (b-g) Pore lifetime are indicated by colors and isolines in different parameter planes represented by dashed lines in panel (a). Symbol are triangles if $\alpha_s$ is less than $0.1$ or circles otherwise. The correspondence between the change of symbol type and the color indicates that $\alpha_s$ being less or greater than $0.1$ is a good criteria for determining if the pore is short or long-lived respectively.}
  \label{fig:phases}
\end{figure*}

\begin{figure}[tb]
\centering
  \includegraphics[scale=1]{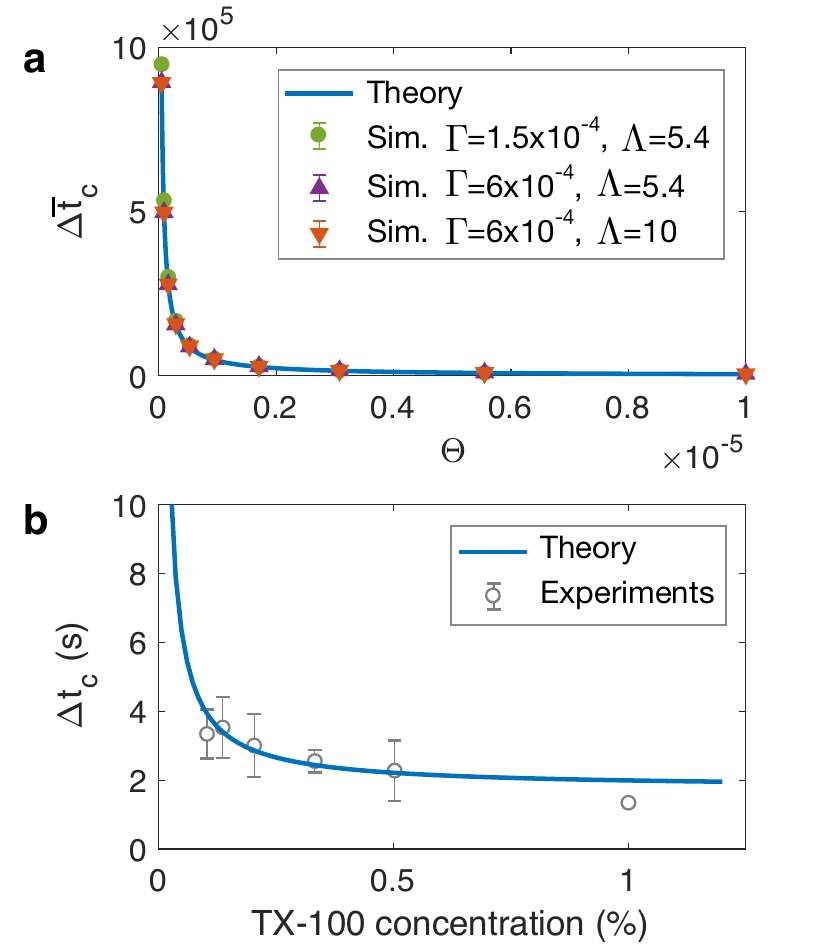}
  \caption{The cycle period is an inverse function of $\Theta$, and is independent of $\Gamma$ and $\Theta$. (a) Dimensionless cycle period as a function of $\Theta$. The theoretical expression (Eq.~\ref{eq:cyc_per_dimless}) is in excelent agreement with simulations (averages and standard deviations of the first three cycle periods, error bars are smaller than symbols). In all simulations $\Phi=4.4\times 10^{-11}$. (b) The theoretical expression of dimensional cycle period fits experimental measures of DOPC vesicles exposed to TX-100 surfactant (data from \cite{hamada2009}) for the solubilization rate defined as $k=k_0 c /(K+c)$ with $c$ the TX-100 volumic concentration, $k_0=5.39 \times 10^{-3}$ s$^{-1}$ and $K=12.6 \%$.}
  \label{fig:cyc_per}
\end{figure}

\subsection{Model validation of short and long-lived pore dynamics}

We first evaluate the ability of this model to reproduce the two regimes of short and long-lived pore dynamics in lipid vesicles exposed to surfactant. To do so, we use model parameters for POPC lipid membranes \cite{chabanon2017}, and adjust the value of three parameters to account for the presence of surfactant: the solubilization rate $k$ \cite{nomura2001, hamada2009, hamada2012}, the pore line tension $\gamma$ \cite{nomura2001, karatekin2003a, puech2003}, and the stretch modulus $\kappa$.

To adjust the value of these parameters, we choose the pore dynamics reported in the experimental study by \citet{hamada2009}, where DOPC vesicles were exposed to various concentrations of Triton X-100 (TX-100) surfactant. 
We solve the two coupled equations for the vesicle radius (Eq.~\eqref{eq:dRdt_dimless}) and pore radius (Eq.~\eqref{eq:drdt_dimless}) with the expression for the reference vesicle radius (Eq.~\eqref{eq:R0}). Numerical results for pore radii of short and long-lived pores are presented in Fig.~\ref{fig:comp} for the parameters shown in Table~\ref{tab:para}. The model predictions are in good agreement with the experimental measurements from \citet{hamada2009}. Our results confirm that short-lived pores are obtained at low solubilization kinetics, corresponding to small surfactant concentration, while long-lived pores occur at fast solubilization rates, equivalent to high surfactant concentrations. 

The semi-quantitative agreement between our model results and the experimental data confirms that: (i) the solubilization rate $k$ is larger with increased concentration of surfactant \cite{nomura2001, hamada2009, hamada2012}; (ii) the pore line tension $\gamma$ is decreased from a typical value of 15 pN in the absence of surfactant \cite{portet2010}, to 1.2 pN and 0.3 pN for low and high concentrations of surfactant respectively, in agreement with experimental measurements of line tension in DOPC  lipid vesicles exposed to Tween 20 surfactant \cite{karatekin2003a, puech2003}; (iii) the value of the stretch modulus $\kappa=$ 0.2$\times 10^{-4}$ N/m is one order of magnitude lower than the one reported for bursting vesicles in the absence of surfactant \cite{chabanon2017, su2018}. Although this value is significantly lower than the elastic stretch modulus of lipid membranes ($\sim$ 0.2 N/m \cite{evans1990}), in this modeling approach $\kappa$ should be regarded as an effective stretch modulus that accounts for the elastic membrane response as well as the unfolding of submicroscopic wrinkles produced by the sudden pore opening \cite{chabanon2017}. The value of this effective modulus has been reported in the absence of surfactant to be 2$\times 10^{-3}$ N/m in the case of pure POPC vesicles \cite{chabanon2017}, and 6$\times 10^{-3}$ N/m in the case of POPC/SM/Ch ternary lipid mixture \cite{su2018}. Here, our results suggest that the presence of surfactant lowers $\kappa$ an order of magnitude independently of the surfactant concentration.

For completeness, we also report the dynamics of the vesicle radius in short and long-lived pores regimes in Figure~\ref{figS:dynamics}, showing qualitative agreement with the stepwise and continuous shrinkage described in the literature \cite{nomura2001, hamada2009, hamada2012}.

\subsection{Influence of dimensionless parameters on the pore dynamics}

Next, we aim to identify the relevant physical processes leading to a short or long-lived pore. Based on our simulations of the pore dynamics, we asked how does the presence of surfactant affect (i) the pore life time, (ii) the maximum pore radius, (iii) the observed (average) pore radius, and (iv) the pore closure dynamics. To answer these questions, we defined four pore metrics represented in Fig.~\ref{fig:MTheta}(a, b). The most intuitive metrics are the pore life time $\Delta \bar{t}_p$, and the maximum pore radius $\bar{r}_\text{max}$.  Our preliminary results of pore dynamics suggest that the pore closure can typically be divided into three distinct phases (Fig.~\ref{fig:MTheta}(a, b)): (i) a short and fast quasi-linear decrease of radius, (ii) a slower and possibly longer closure phase in the case of long-lived pores, and (ii) the final closure. Based on this observation, we defined the observed radius $\bar{r}_\text{obs}$ as the mean radius during phase (ii) of the pore closure. Finally, to characterize the pore closure dynamics, we defined $\alpha_s$ as the ratio between the slopes of the slow closing phase (ii) and fast closing phase (i) (see Fig.~\ref{fig:MTheta}(a, b)). All theses parameters were computed automatically following the algorithm given in Section~\ref{S_metrics} of the Supplementary Material.

The most critical effects of surfactant on lipid membrane are the lost of surface area by solubilization, and the lowering of the pore line tension. Therefore, we first study the effect of the dimensionless parameters $\Theta$ and $\Gamma$ on the pore dynamics. Based on the characteristic parameters from Fig.~\ref{fig:comp}, we set $\Lambda=5.4$, $\Phi=4.4\times 10^{-11}$, and choose three values for $\Gamma=1.5\times 10^{-4}$, $4\times 10^{-4}$, and $6\times 10^{-4}$. The resulting pore life time, maximum pore radius, observed pore radius, and slope ratio are shown in Fig.~\ref{fig:MTheta}(c-e) for values of $\Theta$ ranging from $5\times 10^{-8}$ to $10^{-5}$. Short and long-lived pore regimes are clearly identified from the pore lifetime: for small values of $\Theta$ corresponding to a slow solubilization rate, the first pore to open has a short life time, while large values of $\Theta$ lead to long first pore life time (Fig.~\ref{fig:MTheta}(c)). Surprisingly, we observe a sharp transition between the two regimes of short and long-lived pores, as seen from a sudden increase of about two orders of magnitude in pore life time. This transition from short to long-lived pore can also be seen from the drop of observed pore radius (Fig.~\ref{fig:MTheta}(d)) and slope difference (Fig.~\ref{fig:MTheta}(e)). However, the maximum pore radius is not affected by $\Theta$ nor $\Gamma$  (Figs.~\ref{fig:MTheta}(d) and \ref{figS:Gamma}). The value of the slope ratio $\alpha_s$ is close to one in the short-lived pore regime, and drops two order of magnitude in the long-lived pore regime (Fig.~\ref{fig:MTheta}(e)). Then for increasing values of $\Theta$, the slope ratio progressively increases by one order of magnitude. Note that the plateau shown by $\alpha_s$ at large $\Theta$ values and $\Gamma=1.5\times 10^{-4}$ corresponds to limitations in the fitting procedure to determine the slopes of the two closure phases due to the extreme values of the model parameters (see Fig.~\ref{figS:fits} for the influence of $\Theta$ on the fitting performance of various pore dynamics).

While the pore lifetimes of short-lived pores are weakly dependent on $\Theta$, all pore lifetimes are longer for smaller values of $\Gamma$, or equivalently, for smaller line tensions. Importantly, decreasing $\Gamma$ induces long-lived pores to occur for slower solubilization rates, as seen in the shift of the transition from short to long-lived pores toward lower values of $\Theta$ (see also Fig.~\ref{figS:Gamma} for the influence of $\Gamma$ on the pore metrics).

Next, we study how $\Lambda$, the ratio between the viscous dissipation of the solution and membrane, affects the pore dynamics. Fig.~\ref{fig:Lambda} shows the four metrics of the first pore for $\Lambda=1$, 5.4 and 10 as a function of $\Theta$, with other parameters fixed to $\Gamma=6\times 10^{-4}$ and $\Phi=4.4\times 10^{-11}$. We observe that increasing the value of $\Lambda$ shifts the value of $\Theta$ at which the pore transitions from short to long-lived pores. Furthermore, the higher the value of $\Lambda$, the longer is the pore lifetime. This effect should be considered with caution as the membrane viscosity appear both in $\Lambda$ and the characteristic time $\tau$. Finally, and in contrast to the effect of $\Gamma$, the maximum pore radius increases with $\Lambda$, to approach a maximum value around $\bar{r}_\text{max}=0.55$ (see also Fig.~\ref{figS:Lambda} for the influence of $\Gamma$ on the pore metrics).

Finally, we investigate the influence of $\Phi$ on the pore dynamics. We find that the overall vesicle and pore dynamics are not affected by values of $\Phi$ below 10$^{-6}$. Larger values correspond to unphysical membrane permeability values (results not shown). Characteristic values of $\Phi$ are at least six orders of magnitude smaller than the other dimensionless parameters, suggesting that solvent permeation through the membrane has a negligible effect on the dynamics of lipid vesicles exposed to surfactants.

\subsection{Phase diagram of short and long-lived pore dynamics}

The results above emphasize that the three critical dimensionless parameters determining the first pore dynamics are $\Theta$, $\Gamma$, and $\Lambda$. This motivates a systematic exploration of how the combinations of these parameters lead to short or long-lived pores. 

The results presented in Figures~\ref{fig:MTheta} and \ref{fig:Lambda} suggest that the two most relevant metrics allowing to characterize short and long-lived pores are the pore lifetime $\Delta \bar{t}_p$ and the slope ratio $\alpha_s$. Therefore we investigate the values of these characteristics for physical range of $\Theta$, $\Gamma$, and $\Lambda$ in terms of short or long-lived pore. Figure~\ref{fig:phases}(a) shows isocontours of the pore lifetime in the $(\Theta, \Gamma, \Lambda)$ parameter space. These results confirm our previous observations that the pore life time is either above $10^{5}$ for long-live pores, or below $10^{4}$ for short-lived pores, with a sharp transition from one regime to the other (very few values of $\Delta \bar{t}_p$ between $10^{4}$ and $10^{5}$). Longer pore lifetimes $\Delta \bar{t}_p\ge 10^{5}$ are obtained for large values of $\Theta$ and $\Lambda$, and small values of $\Gamma$. The transition from short to long-lived pore corresponds to a plane in the logarithmic parameter space. This can be further seen in Figures~\ref{fig:phases}(b-g) where the pore life time within the parameter planes represented by dashed lines in Figure~\ref{fig:phases}(a) are shown along with the isocontours. In addition, circular symbols indicate that the slope ratio $\alpha_s$ is above the value $0.1$, characteristic of LLP, while triangular symbols indicate $\alpha_s<0.1$. For most values of the parameters investigated, the agreement between color scheme (pore life time) and symbol (slope ratio above or below $0.1$) suggests that the value of $\alpha_s$ is a good indicator of the pore dynamic regime, and allows us to discriminate between short and long-lived pore regimes. It should be noted that for high values of $\Theta$, our methodology measures $\alpha<0.1$ despite long pore lifetimes. This actually arises from the limitations of the fitting procedure. Examples are presented in Fig.~\ref{figS:fits}.


\subsection{The cycle period between short-lived pores is an inverse function of the solubilization rate}

Finally, we characterize how the solubilization kinetics influences the sequence of pore formation. We define the dimensionless cycle period $\Delta \bar{t}_c = \Delta t_c / \tau$ as the dimensionless time between two successive pore closings, starting at the end of the first pore opening. 

First, we approximate analytically the cycle periods based on Eqs.~\eqref{eq:dRdt_dimless} and \eqref{eq:dR0dt_dimless}. The detailed derivation is presented in Section~\ref{S_cyc_period} of the Supplementary Material, resulting in the following expression for the cycle period
\begin{equation} \label{eq:cyc_per_dimless}
\Delta \bar{t}_c = \frac{\epsilon^*}{\epsilon^*+1} \frac{1}{2\Theta} ,
\end{equation}
where $\epsilon^*$ is the observed lytic area strain \cite{chabanon2017}. Remarkably, the cycle period is independent of $\Gamma$ and $\Lambda$. The cycle period decreases with the solubilization kinetic parameter, in agrement with previously reported experimental observations \cite{hamada2009}.
The analytic cycle period (Eq.~\eqref{eq:cyc_per_dimless}) is plotted in Fig.~\ref{fig:cyc_per}(a) assuming $\epsilon^*=0.1$ \cite{chabanon2017}, together with computed values of $\Delta\bar{t}_c$ from numerical simulations with various parameters. The theoretical expression for the cycle period is in excellent agreement with all numerical results, confirming the independence of $\Delta\bar{t}_c$ on $\Gamma$ and $\Lambda$.


In order to compare the cycle periods with experimental data, we assume that the solubilization rate is dependent on the surfactant concentration $c$ such as $k=k_0 c / (K + c)$, where $k_0$ and $K$ are the surfactant specific parameters to be determined. This dependence assumes a saturation of the effects of the surfactant on the solubilization rate at high concentrations, as expected. We can now write the dimensional cycle period as a function of the surfactant concentration such as
\begin{equation} \label{eq:cyc_per}
\Delta t_c = \frac{\epsilon^*}{\epsilon^*+1} \frac{ K + c}{k_0 c} .
\end{equation}
In order to determine the surfactant parameters $k_0$ and $K$, we fit Eq.~\ref{eq:cyc_per} to the experimental measures of cycle periods of DOPC vesicles exposed to various concentration of TX-100 surfactant reported in \cite{hamada2009}. The result, presented in Fig.~\ref{fig:cyc_per}(b), yields $k_0=5.39 \times 10^{-3}$ s$^{-1}$ and $K=12.6 \%$ (goodness of fit $R^2=0.78$), showing a good agreement between Eq.~\eqref{eq:cyc_per} and the experimental cycle periods.

\section{Discussion}

Although equilibrium shapes of membrane systems are relatively well understood \cite{kas1991, lipowsky1991, seifert1997, lipowsky2014, ho2016, hassinger2017, chabanon2017a}, the description of their out-of-equilibrium behavior remains a major challenge. 
In this article, we propose a quantitative physical model for the out-of-equilibrium dynamics of lipid vesicle induced by surfactant. We show how the dynamics of the first microscopic pore can be either short or long-lived, depending on the surfactant and membrane properties. The driving mechanism for this behavior is the solubilization of the lipid bilayer, which induces an area reduction of the vesicle at almost constant volume. The progressive reduction of the area to volume ratio produces an increase in membrane tension, eventually leading to membrane rupture, and the opening of a large micrometer-sized pore. Interestingly, two possible scenarios occur at this point (Fig.~\ref{fig:solu}(a)): either the pore closes in about a second after opening (short-lived pore), or the pore stays open for a long time, typically between ten seconds and a minute, before closing (long-lived pore). Then, as area reduction of the vesicle continues, subsequent series of short-lived pores occur independently of the first pore dynamics, until total solubilization of the lipid vesicle is completed.

We propose a model for pore and vesicle dynamics that recapitulates the two first pore regimes as well as the subsequent cycle dynamics. The key component of the model is the solubilization rate $k$, which determines how fast the membrane tension $\sigma$ builds up by reducing the reference membrane area $A_0$ (see Fig.~\ref{fig:solu}(c) and Eq.~\eqref{eq:sigma}). Indeed, the pore opening and closing is determined by the balance between membrane tension, which tends to open the pore, and pore line tension $\gamma$, which tends to close the pore. At slow solubilization rates, the membrane tension induced by area reduction does not build fast enough to prevent the pore from closing by line tension. However, a high solubilization rate increases membrane tension fast enough to prevent the pore closing, leading to a long-lived pore. For a vesicle with an open pore, membrane tension is released by the leak-out through the open pore, driven by the Laplace pressure. Since Laplace pressure is an inverse function of the vesicle radius, the leak-out rate is faster for small vesicles, and allows a faster reduction of the membrane tension as the vesicle decreases in size. As a result, vesicles exhibiting long-lived pores eventually reseal as their size decreases, and show only subsequent short-lived pores \cite{hamada2009}.


Although short and long-lived pores in lipid vesicles exposed to surfactant have been observed experimentally \cite{nomura2001, hamada2009, tomita2011, hamada2012}, the physical understanding of this phenomena has been limited \cite{kaga2006, hamada2009}. Here we propose a model that can semi-quantitatively reproduce the pore dynamics reported in experimental studies, allowing us to investigate the effect of the lipid bilayer properties and surfactant solution on the vesicle and pore dynamics.

In order to quantitatively study the first pore dynamics as a function of the system parameters, we defined four pore metrics (Fig.~\ref{fig:MTheta}(a, b)). Surprisingly, we find that, as we increase the solubilization kinetics, the system transitions abruptly from short to long-lived pore regime. Furthermore, we show that this transition can be modulated by two dimensionless numbers. First, we discuss the role of $\Gamma$ (defined in Eq.~\eqref{eq:Gamma}), which represents the ratio between pore line tension and the membrane stretch modulus. We show that small values of  $\Gamma$ -- corresponding to small line tension and high membrane tension -- facilitates the occurrence of long-lived pores (Fig.~\ref{fig:MTheta}(c-e)). Second, we discuss the role of $\Lambda$ (defined in Eq.~\eqref{eq:Lambda}), which represents the ratio between solution and membrane viscosity. We find that long-lived pores are favored by large values of $\Lambda$ (Fig.~\ref{fig:Lambda}), \textit{i.e.} by solutions of high viscosity that slow down the leak-out and therefore membrane tension relaxation. By systematically computing the pore metrics for a large number of combinations of the relevant dimensionless parameters, we show that the regimes of short and long-lived pores are separated by a plane in the  $(\Theta, \Gamma, \Lambda)$ logarithmic parameter space. Finally, we turn our attention to the dynamics of the subsequent short-lived pores. We show, both numerically and theoretically, that the cycle period depends only on the solubilization rate and the membrane lytic strain (Fig.~\ref{fig:cyc_per}(a)). Our theoretical expression of the cycle period is in excellent agreement with experimental data reported in the literature (Fig.~\ref{fig:cyc_per}(b)).

It should be noted that, the reason why our model predictions are compared with experimental data from DOPC/TX-100 lipid/surfactant systems only, is because, to our knowledge, no other quantitative measurement of pore dynamics in lipid vesicles exposed to surfactant are available in the literature \cite{hamada2009, hamada2012}. However, a large number of experimental studies report qualitative observation of cyclic pore opening in lipid vesicles based on various combinations of POPC, DOPC, DMPC, PG, DMPG, PA, DMPA, DMDAP and DMTAP, exposed to different concentrations of TX-100, Tween 20, sodium cholate, octyl glucoside, polyoxyethylene 8 lauryl ether, CHAPS hydrate, Sulfobetaine 3-14, hexadecyl pyridinium chloride, hexadecyl trimethyl ammonium bromide, ethanol, DL-pyrrolidonecarboxylic acid salt, and polyoxethylene (caprylate/caprate) glycerides \cite{nomura2001, tomita2011, hamada2009, hamada2012}. Thus, the theoretical framework proposed here provides a foundation for quantitative studies of lipid vesicle solubilization dynamics.


Despite showing semi-quantitative agreement with reported experimental data, and giving insight on the pore dynamics regime as a function of the membrane and surfactant parameters, our model does have its limitations. Vesicle area reduction by surfactant, although the most frequent outcome \cite{nomura2001, karatekin2003, karatekin2003a, tomita2011, hamada2012}, is not the only response of lipid vesicles exposed to surfactants. Depending on the combination of surfactant and membrane composition, the vesicle can exhibit complex topological changes leading to invaginations, fission, formation of multilamellar structures, or complete bursting \cite{nomura2001, tomita2011, hamada2012}. This variability is attributed to various parameters such as the surfactant-membrane affinity, the spontaneous curvature and flip-flop kinetics of the surfactant, the fluidity of the membrane, as well as its compositional heterogeneity \cite{nomura2001, tomita2011, hamada2012, nazari2012}. Further theoretical work and systematic experimental characterization are needed to quantify the respective importance of theses factors on the vesicles fate. 

For most surfactants/lipid bilayer combinations, the initial stage of solubilization is characterized by a destabilization of the vesicle shape with possible invaginations \cite{hamada2009, hamada2012}. It is then in a second stage that area reduction occurs, leading to increased tension, flattening of the membrane fluctuations, and recovery of a spherical shape, eventually followed by cyclic pore opening \cite{hamada2012}. Our model focuses on this second stage, where the vesicle is well approximated by an elastic sphere. It should also be noted that our model does not account for the final stage where the radius of the vesicle reaches the micrometer size and less. Discrete numerical models such as coarse grain modeling are more suited to study solubilization at such scales \cite{noguchi2012}.

And finally, the dynamics of the system we modeled comes from the time evolution of the vesicle geometry. It is possible that the solubilization kinetics affects the physical parameters of the lipid membrane in a time-dependent manner. How fast the pore line tension, membrane viscosity, or stretch modulus are going to be affected by the increase of surfactant surface concentration in the membrane will influence the observed vesicle dynamics. Two limiting cases can be considered:  (i) the surfactant insertion limited regime, where micelle formation is much faster than surfactant insertion. In that case the concentration of surfactant in the membrane increases with time at almost constant membrane surface area, leading to dynamic variations of the membrane properties. (ii) The solubilization limited regime (or slow solubilization regime \cite{lichtenberg2013a}), where the formation of micelles is much slower than the surfactant insertion in the membrane. In that case all the solubilization process occurs when the membrane is saturated in surfactant, allowing for the assumption that the membrane properties are constant. Based on the observations that lipid vesicles exposed to surfactant first exhibit shape destabilization before undergoing area reduction, it is possible that lipid solubilization occurs at a threshold membrane surfactant concentration, triggering the transition from the first to the second regime. While our model lies in the solubilization limited regime, including time-dependent membrane parameters is straight forward, allowing us to represent an intermediate regime where the surfactant insertion and micelle formation occurs on a similar time scale. Yet, few estimates of the solubilization kinetics are available, pointing out the need for more experimental studies on surfactant kinetics \cite{lichtenberg2013a}. We believe that the model presented in the present study can be used as a framework to help characterizing solubilization kinetics of lipid membrane.




Cyclic pore openings in lipid vesicles is not exclusively induced by surfactants. In fact, similar vesicle dynamics have been observed for a variety of external stressors. Swell-burst cycles were first predicted theoretically for small unilamellar vesicles in hypotonic conditions \cite{koslov1984}, and later observed in artificial giant unilamellar vesicles of various lipid composition \cite{peterlin2008, oglecka2014, ho2016, chabanon2017, su2018}. In that case, the increase in membrane tension is driven by the osmotic influx of water through the membrane, which produces a cyclic series of swelling and bursting of the vesicle. Likewise, lipid vesicles undergoing photooxidation have shown to exhibit series of pore openings very similar to those observed with surfactant, except that swelling phases occur in-between bursting events \cite{karatekin2003a, sankhagowit2014}. This behavior is attributed to both area reduction and osmotic imbalance due to the release of photo-oxidative products \cite{mabrouk2010, peyret2017}. Such behavior was also reported in light activated polymerosomes \cite{mabrouk2010, peyret2017, albertsen2017}. Taken together, these observations suggest that cyclic opening of large pores is a general mechanism allowing cell-sized vesicle to maintain their integrity against a variety of environmental attacks. By proposing a quantitative biophysical model of pore dynamics in lipid vesicles exposed to surfactant, we undertake an essential step toward a better understanding of the fundamental mechanisms allowing cells to endure constantly changing environments, and provide an important theoretical tool to aid the design of vesicle based drug delivery systems.




\section*{Acknowledgments}

This work was supported in part by the AFOSR FA9550-15-1-0124 award, the ARO W911NF-16-1-041 award, and the ONR N00014-17-1-2628 award to P.R. The authors are grateful to Miriam Bell for critical reading.

\section*{References}

\bibliographystyle{elsarticle-harv}
\bibliography{Surfactant_Oxidation}


\pagebreak


\noindent{\Large \bf Supporting Material for:}\\

\noindent{\LARGE \bf Solubilization kinetics determines the pulsatory dynamics \\of lipid vesicles exposed to surfactant} 
\vspace{0.6cm}  \\
\noindent{\large \bf Morgan Chabanon and Padmini Rangamani }  \vspace{0.1cm} \\
\noindent \small{Department of Mechanical and Aerospace Engineering, University of California San Diego, La Jolla, California.}


\setcounter{page}{1}
\setcounter{section}{0}
\setcounter{figure}{0}
\counterwithin*{equation}{section}
\renewcommand{\thesection}{S\arabic{section}} 
\renewcommand{\theequation}{\thesection.\arabic{equation}} 
\renewcommand{\thefigure}{S\arabic{figure}} 

\section{Characterization of the pore dynamics} \label{S_metrics}

\begin{figure*}[tbp]
\centering
\includegraphics[scale=1]{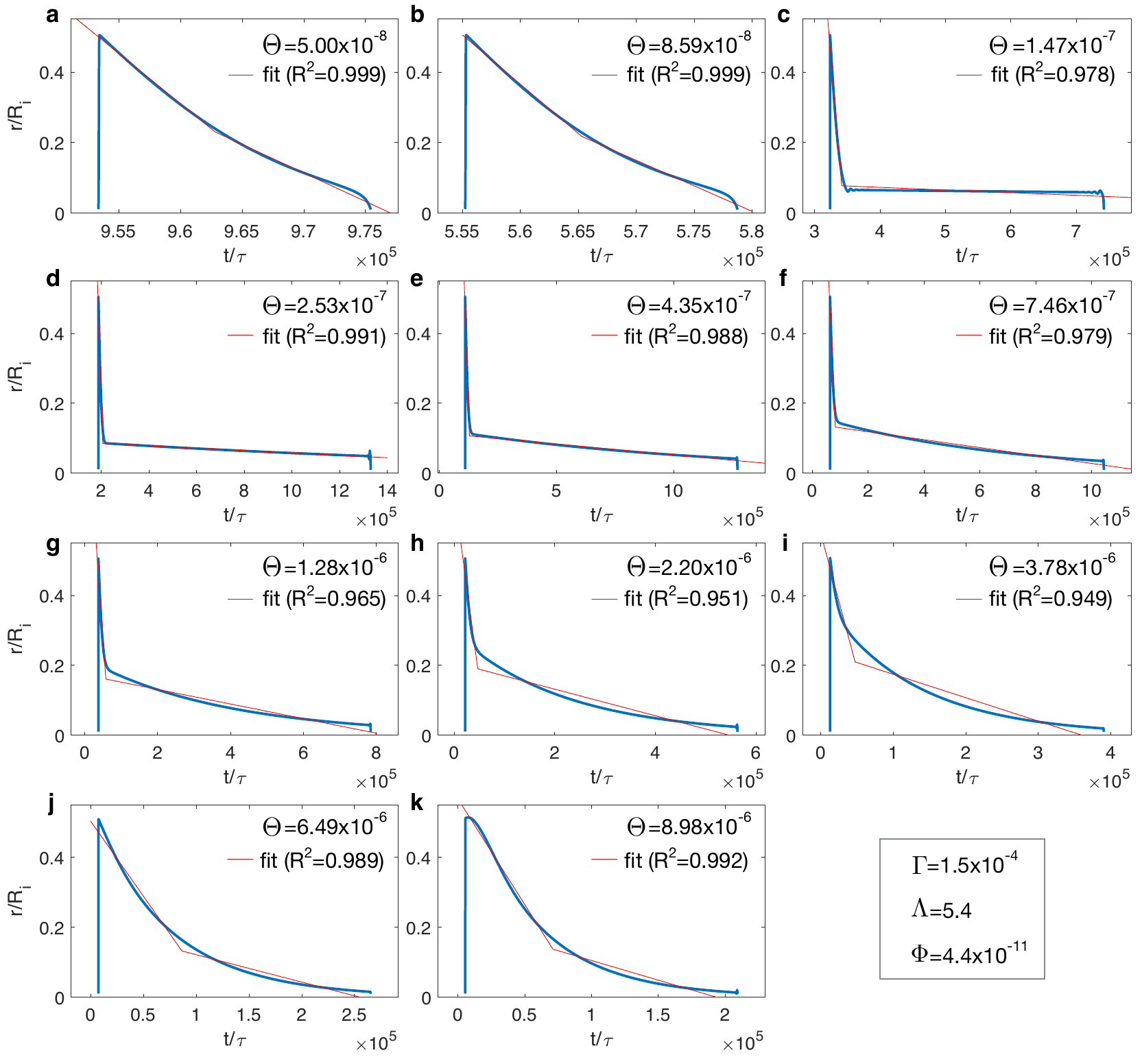}
\caption{Typical pore dynamics and corresponding fit for increasing values of $\Theta$. The overall fitting procedures yields excellent results, except for higher values of $\Theta$ where the pore closure dynamics progressively deviates from two intersecting straight lines.}
\label{figS:fits}
\end{figure*}

All parameters are evaluated during the first pore opening, defined as $\bar{r}>0.01$. 
The pore life time is defined as $\Delta \bar{t}_p=\bar{t}_\text{close} - \bar{t}_\text{open}$, and we define $\bar{r}_\text{max}= \text{max}(\bar{r})$.
We then fit the following equation to the radius between $\bar{r}_\text{max}$ and the last inflection point of the pore radius:
\begin{equation}
y=(a_0 + a_1 x_0) + \frac{a_1 + a_2}{2}(x-x_0) 
+ \frac{a_2 - a_1}{2}(x-x_0) \frac{x-x_0}{\mid x-x_0 \mid} .
\end{equation}
This equation is one of two lines intersecting at $x_0$ such that $y(x\le x_0) = a_0 + a_1 x$ and $y(x\ge x_0) = a_0 + (a_1-a_2) x_0 + a_2 x$ (see Fig.~\ref{fig:MTheta}(a, b)). The fitted value of $x_0$ gives us the transition time between the fast closure phase and the slow closure phase. We further define the observed radius as the mean radius of the slow closure phase $\bar{r}_\text{obs}=\langle \bar{r} \rangle$, and the slope ratio as $\alpha_s=a_2/a_1$. Pore dynamics and corresponding fits are shown in Fig.~\ref{figS:fits}.

\section{Analytical expression for the cycle period} \label{S_cyc_period}

Here we derive an analytical expression for the time between two successive short-lived pores in lipid vesicles in the presence of surfactant.

The areal membrane strain of a spherical vesicle without pore is given by
\begin{equation}
\epsilon = \left( \frac{\bar{R}}{\bar{R_0}} \right)^2 - 1 .
\end{equation}
It follows that the dimensionless strain rate is
\begin{equation} \label{eq:Sstrainrate}
\frac{d\epsilon}{d\bar{t}} = \frac{2 \bar{R}}{\bar{R}_0^2} \left( \frac{d\bar{R}}{d\bar{t}} - \frac{\bar{R}}{\bar{R}_0  } \frac{d\bar{R}_0}{d\bar{t}} \right) .
\end{equation}
Introducing Eqs.~\eqref{eq:dRdt_dimless} (for a closed pore $\bar{r}=0$) and \eqref{eq:dR0dt_dimless} into Eq.~\eqref{eq:Sstrainrate}, we have
\begin{equation} \label{eq:Sstrainrate2}
\frac{d\epsilon}{d\bar{t}} = \frac{2 \bar{R}}{\bar{R}_0^2} \left( - \frac{\Phi}{\Lambda} \frac{4\pi \epsilon}{\bar{R}} + \bar{R} \Theta  \right) .
\end{equation}
During the cyclic regime of short-lived pores, the cycle period is approximately equal to the time the vesicle needs to reach the lytic strain. Therefore we can write
\begin{equation}
\Delta \bar{t}_c = \epsilon^* \left( \frac{d\epsilon}{d\bar{t}} \right)^{-1} ,
\end{equation}
where $\epsilon^*$ is the observed critical strain. It follows that
\begin{equation}
\Delta \bar{t}_c = \frac{\bar{R}_0^2}{ \bar{R}^2 }  \frac{\epsilon^*}{2\Theta}  \left( - \frac{\Phi}{\Lambda\Theta} \frac{4\pi \epsilon^*}{\bar{R}^2}  +1 \right)^{-1} .
\end{equation}
Examining the order of magnitude of the first term in the parentheses, we have 
\begin{equation}
\frac{\Phi}{\Lambda\Theta} \frac{4\pi \epsilon^*}{\bar{R}^2}  \sim 10^{-6} - 10^{-4} \ll 1 ,
\end{equation}
allowing us to express the cycle period as
\begin{equation}
\Delta \bar{t}_c = \frac{\bar{R}_0^2}{ \bar{R}^2 }  \frac{\epsilon^*}{2\Theta} .
\end{equation}
By noting that when $\epsilon=\epsilon^*$, $(\bar{R}/\bar{R}_0)^2=\epsilon^*+1$, we obtain that
\begin{equation} \label{eq:Scyc_per}
\Delta \bar{t}_c = \frac{\epsilon^*}{\epsilon^*+1} \frac{1}{2\Theta} .
\end{equation}
Remarkably, this expression of the cycle period is independent of $\Gamma$ and $\Lambda$. The dimensional form of the cycle period is
\begin{equation} \label{eq:Scyc_per2}
\Delta t_c = \frac{\epsilon^*}{\epsilon^*+1} \frac{1}{k} .
\end{equation}
Thus we show that cycle period decreases with the solubilization kinetic parameter, in agreement with previously reported experimental observations \cite{hamada2009}. 

The dimensionless cycle period (Eq.~\eqref{eq:Scyc_per}) is plotted in Fig. \ref{fig:cyc_per}(a) for $\epsilon^*=0.1$ \cite{chabanon2017}, yielding an excellent agreement with numerical results.
In order to compare the cycle periods with experimental measurements, we assume that the solubilization rate is dependent on the surfactant concentration $c$ such as $k=k_0 c / (K + c)$, where $k_0$ and $K$ are the surfactant specific coefficients to be determined. To do so, we fit the resulting dimensional equation
\begin{equation} \label{eq:Scyc_per3}
\Delta t_c = \frac{\epsilon^*}{\epsilon^*+1} \frac{ K + c}{k_0 c} ,
\end{equation}
with $\epsilon^*=0.1$ to the experimental measurements of cycle period of DOPC vesicles exposed to various concentrations of TX-100 surfactant presented in \cite{hamada2009}. A plot of Eq.~\eqref{eq:Scyc_per3} is shown in Fig.~\ref{fig:cyc_per}(b) for the obtained values of $k_0=5.39 \times 10^{-3}$ s$^{-1}$ and $K=12.6 \%$.

\section{Supplementary figures}

\begin{figure*}[hp]
\centering
\includegraphics[scale=1]{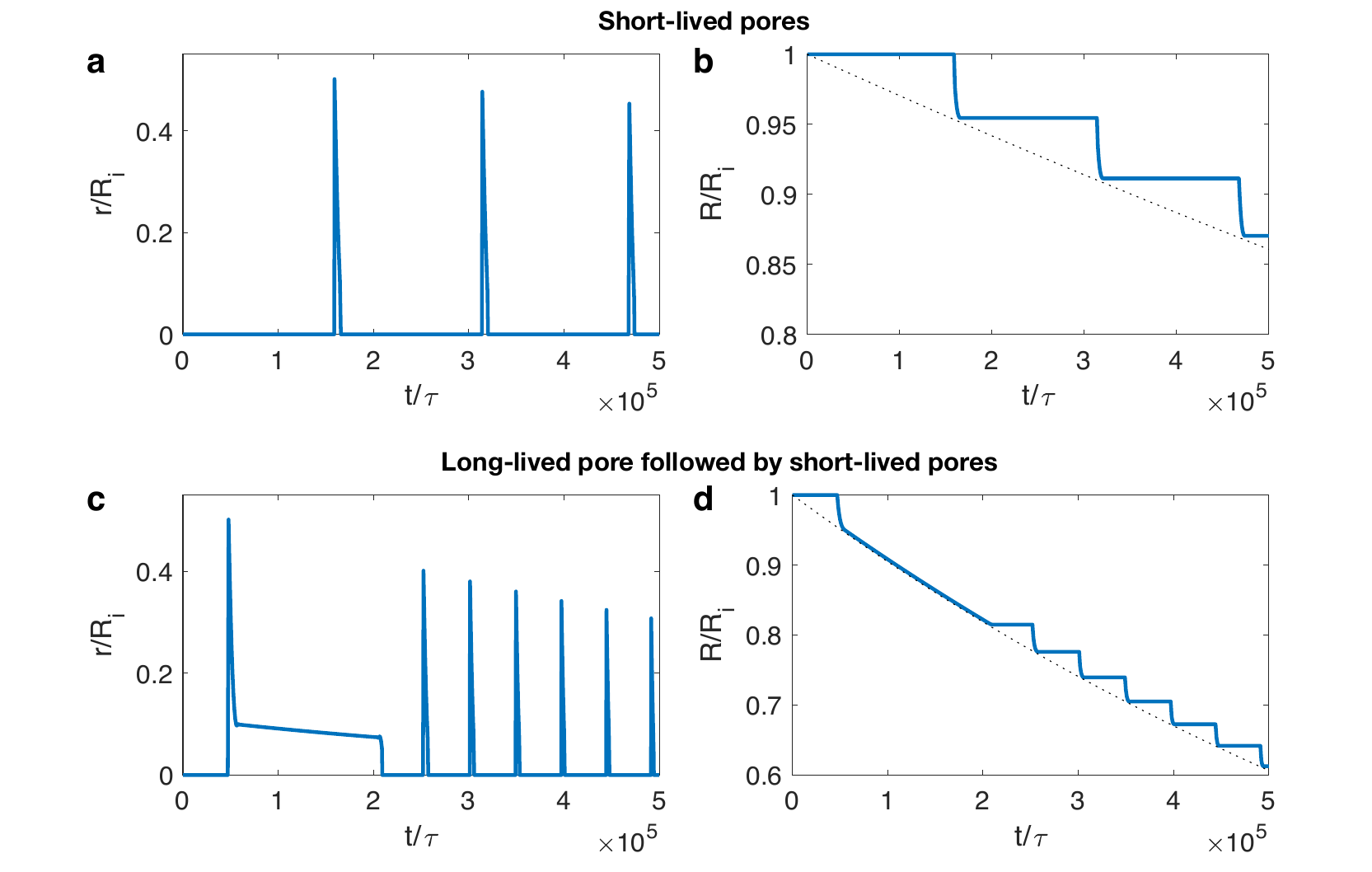}
\caption{Typical pore and vesicle dynamics in short-lived (a, b) and long-lived (c, d) regimes. Dimensionless pore radii (a, c) and vesicle radii (b, d) as a function of dimensionless time. Black dashed lines in panels (b, d) represent the dimensionless reference radius $R_0/R_i$. (a, b) $\Theta = 3\times 10^{-7}$. (c, d) $\Theta = 1\times 10^{-6}$. Other parameters are constant in all panels: $\Gamma = 1.5\times 10^{-4}$, $\Lambda=5.4$, and $\Phi = 3\times 10^{-7}$.}
\label{figS:dynamics}
\end{figure*}

\begin{figure*}[tbp]
\centering
\includegraphics[scale=1]{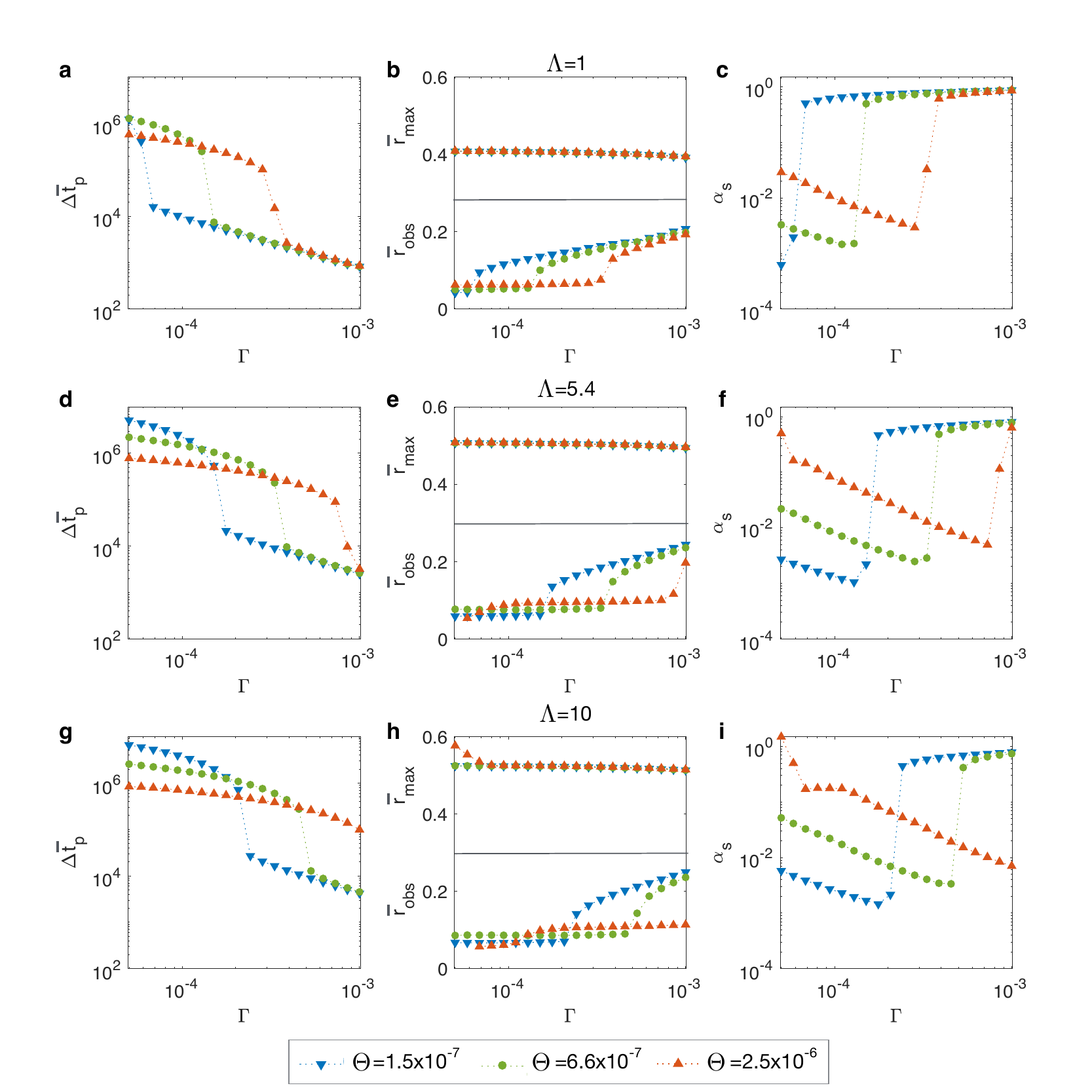}
\caption{Pore metrics as a function of $\Gamma$ for various values of $\Theta$ and $\Lambda$. (a-c) $\Lambda = 1$. (d-f) $\Lambda = 5.4$. (g-i) $\Lambda = 10$. Large values of $\Gamma$ promote short-lived pores, while large values of $\Theta$ and $\Lambda$ promote long-lived pores. In all cases, $\Phi=4.4\times 10^{-11}$. Pore metrics are defined in Fig.~\ref{fig:MTheta}(a, b).}
\label{figS:Gamma}
\end{figure*}

\begin{figure*}[tbp]
\centering
\includegraphics[scale=1]{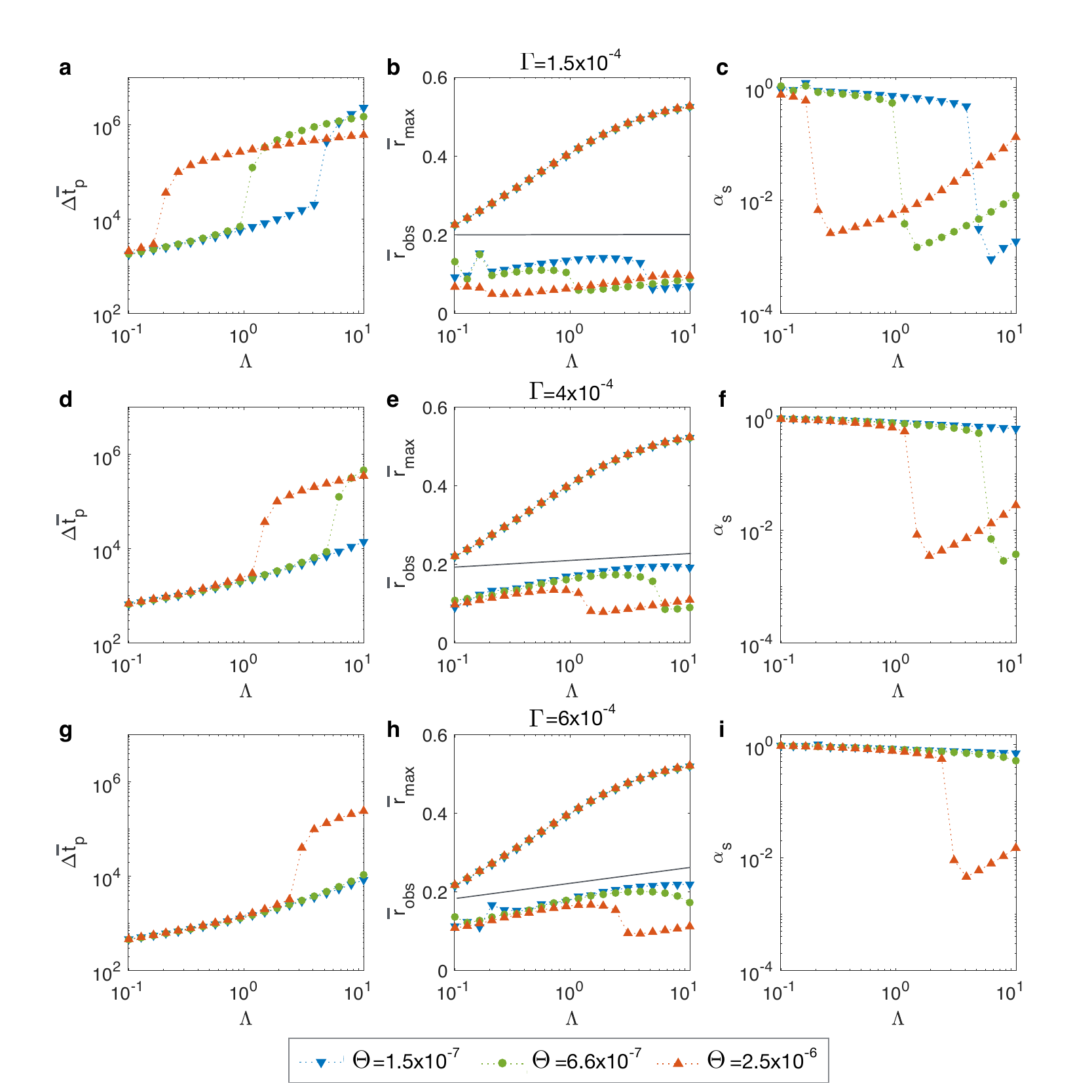}
\caption{Pore metrics as a function of $\Lambda$ for various values of $\Theta$ and $\Gamma$. (a-c) $\Gamma = 1.5\times 10^{-4}$. (d-f) $\Gamma = 4\times 10^{-4}$. (g-i) $\Gamma = 6\times 10^{-4}$ The maximum pore radius is only influenced by the value of $\Lambda$.  In all cases, $\Phi=4.4\times 10^{-11}$. Pore metrics are defined in Fig.~\ref{fig:MTheta}(a, b).}
\label{figS:Lambda}
\end{figure*}

\end{document}